\documentclass[prl, twocolumn, floatfix,showpacs,superscriptaddress]{revtex4-1}

\usepackage{graphicx}
\usepackage{bm}
\usepackage{enumerate}
\usepackage[latin1]{inputenc}
\usepackage{color}
\usepackage[normalem]{ulem} %emphasize weiterhin kursiv

\begin{document}

\preprint{APS/123-QED}

\title{Optical Spin Noise of a Single Hole Spin Localized in an (InGa)As Quantum Dot\\
}

\author{Ramin Dahbashi}
% \email{dahbashi@nano.uni-hannover.de}
\author{Jens H\"ubner}
\email{jhuebner@nano.uni-hannover.de}
\author{Fabian Berski}
\affiliation{Institut f{\"u}r Festk\"orperphysik, Leibniz Universit\"at Hannover, Appelstr.~2, D-30167 Hannover, Germany}
\author{Klaus Pierz}
% \email{PTB}
\affiliation{Physikalisch Technische Bundesanstalt, Bundesallee 100, D-38116 Braunschweig, Germany}
\author{Michael Oestreich}
\email{oest@nano.uni-hannover.de}
\affiliation{Institut f{\"u}r Festk\"orperphysik, Leibniz Universit\"at Hannover, Appelstr.~2, D-30167 Hannover, Germany}
\date{\today}

\begin{abstract}
We advance spin noise spectroscopy to the ultimate limit of single spin detection. This technique enables the measurement of the spin dynamic of a single heavy hole localized in a flat (InGa)As quantum dot.
% \tcr{Magnetic field and light intensity dependent studies reveal even at low magnetic fields a strong magnetic field dependence of the longitudinal heavy hole spin relaxation rate and an % extremely long $T_1$ of $\ge 180~\mu$s at 31~mT and 5~K.}
Magnetic field and light intensity dependent studies reveal even at low magnetic fields a strong magnetic field dependence 
% of $B_z^{-3/2}$ 
of the longitudinal heavy hole spin relaxation time with an extremely long $T_1$ of $\ge 180~\mu$s at 31~mT and 5~K.
The wavelength dependence of the spin noise power discloses for finite light intensities an inhomogeneous single quantum dot spin noise spectrum which is explained by charge fluctuations in the direct neighborhood of the quantum dot. The charge fluctuations are corroborated by the distinct intensity dependence of the effective spin relaxation rate.
%The charge fluctuations are corroborated by the distinct intensity dependence of the full width at half maximum of the spin noise spectrum.
\end{abstract}

\pacs{72.25.Rb, 72.70.+m, 78.67.Hc, 85.75.-d}
\maketitle

Optical spin noise spectroscopy (SNS) is in principle a nondestructive measurement technique which has been transferred from quantum optics to semiconductor physics in 2005 \cite{Oestreich2005a}. The technique exploits the ever present random fluctuations of spin polarization at thermal equilibrium which are detected by optical Faraday rotation and contain according to the fluctuation dissipation theorem the full dynamic of the spin system. Spin noise spectroscopy is potentially suited to study prospective quantum information systems like quantum repeaters, where photon imparted spin entanglement plays a crucial role \cite{Julsgaard2004}, or semiconductor spin systems, where optical excitation demolishes the intrinsic spin dynamic, e.g., by carrier heating, creation of free carriers, or electron hole spin relaxation via the Bir-Aronov-Pikus mechanism \cite{Muller2010a, hubner.pssb.2014}. The first SNS measurements in semiconductors were demonstrated on bulk GaAs where about 10 billion electrons contributed to the spin noise (SN) signal \cite{Oestreich2005a}. Three years later, SNS revealed the intrinsic spin lifetime of electrons in (110) quantum wells at an ensemble of about 170,000 electrons \cite{Muller2008a}. In 2012, two experiments demonstrated SNS on quantum dot (QD) ensembles where the signal resulted from as low as 50 heavy holes \cite{Dahbashi2012,Li2012}. In this publication we push SNS to the ultimate limit and use the technique to study the fragile spin relaxation dynamic of a single heavy hole localized in a single (InGa)As quantum dot. Thus, SNS finds its way into the very active field of optical single spin detection in quantum dots which has been extremely successful, e.g., studying electron and transverse hole spin dynamic and coherent spin control \cite{atature2007, berezovsky2006, degreve2011}.

During the last few years, strongly localized heavy holes in single (InGa)As quantum dots have attracted considerable attention as a new candidate for semiconductor quantum information qubits \cite{Brunner2009,Glazov2012,Sinitsyn2012,Maier2012, Warburton2013}. 
Theory and experiment show that such heavy holes have in comparison to electrons a significantly longer inhomogeneous transverse spin dephasing time $T_2^\ast$ since their \textit{p}-type wave function with vanishing probability density at the nuclei leads to a rather weak, Ising-like hyperfine interaction with the random spin orientation of the nuclear spins \cite{fischer2008, eble2009, crooker2010}. Theory and experiment also consistently show that the longitudinal heavy hole spin relaxation time $T_1$ is long for high longitudinal magnetic fields $B_z$ and decreases in this regime strongly with increasing magnetic field $B_z$ \cite{Heiss2007}. 
On the other hand, studies of $T_1$ at low magnetic fields on the order of the average nuclear magnetic field are difficult, rare, and contradictory but at the same time important for the easy implementation of holes as semiconductor qubits and the investigation of the central spin problem \cite{Szumniak2012,Barnes2012}. 
%\tcr{Theoretical publications} suggest either a negligible \cite{Trif2009} or a very strong \cite{Sinitsyn2012} magnetic field dependence of $T_1$ for very low magnetic fields. \tcr{Other experiments} also either indicate a saturation around $B_z=0$~mT \cite{Gerardot2008b} or show with increasing $B_z$ an increase of $T_1$ by a factor of 2.5 \cite{Fras2012a} or an increase by one order of magnitude which saturates at 10~mT \cite{Li2012}. 
Calculations suggest either a negligible \cite{Trif2009} or a very strong \cite{Sinitsyn2012} magnetic field dependence of $T_1$ for very low magnetic fields. Experiments also either indicate a saturation around $B_z=0$~mT \cite{Gerardot2008b} or show with increasing $B_z$ an increase of $T_1$ by a factor of 2.5 \cite{Fras2012a} or an increase by one order of magnitude which saturates at 10~mT \cite{Li2012}. 
In the following, we show that in fact the magnetic field dependence of $T_1$ at low magnetic fields is huge and that $T_1$ increases monotonically by three orders of magnitude between 0~mT and 31~mT.

\begin{figure}[tbp]
  \centering
  \includegraphics[width=0.95 \columnwidth]{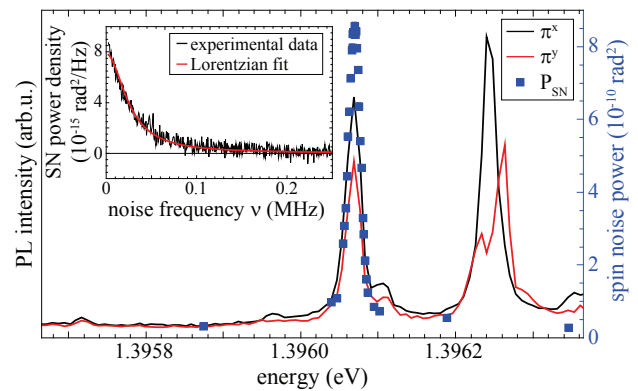}
  \caption{(Color online) The solid black and red lines show the PL spectrum for two orthogonal linear polarizations $(\pi^x, \pi^y)$. The blue squares depict the measured integrated spin noise power $(P_{\text{\tiny SN}})$ which corresponds to the $T_1$ spin noise. The inset shows a typical spin noise spectrum at $B_z=31$~mT after only 12 minutes integration time.
  }
\label{fig:PL}
\end{figure}

The investigated sample is a single layer of self-assembled InAs/GaAs quantum dots grown by molecular beam epitaxy on (100)-oriented GaAs inside the antinode of a $\lambda$-Bragg cavity with 13 and 30 GaAs/AlAs layers for the top and bottom mirror, respectively. The microcavity enables SNS measurements in reflection and enhances the Faraday rotation noise signal without increasing the optical shot noise. The QD emission is shifted to higher energies by vertical QD size reduction and material intermixing during a 6 min growth interruption with a temperature increase up to $590~^\circ\text{C}$ \cite{garcia1998}. 
Across the sample the QD density varies gradually from zero to about 100 dots/$\mu$m$^2$ where a fraction of QDs are filled by a single hole due to a 
% the typical unintentional 
\textit{p}-type background doping density of $10^{14}$ holes/cm$^3$.
The charging of the QDs by holes has been verified by previous SNS ensemble measurements on exactly the same sample but on a sample spot with higher QD density. These measurements show the characteristic heavy hole $T_2^\ast$ time which is about one order of magnitude longer than the corresponding $T_2^\ast$ time of electrons in such QDs \cite{Dahbashi2012}. 
% \tcr{which yield a clear signature of the heavy hole ensemble $T_2^*$ spin dephasing time, i.e., the $B_z$-independent characteristic inhomogeneous spin dephasing time} \cite{Dahbashi2012}.
%The charging of the QDs by holes has been verified before by SNS ensemble measurements on exactly the same sample but on a sample spot with higher QD density \cite{Dahbashi2012}. 
We choose for our SNS measurements a sample region where two quantum dots are in resonance with the cavity and energetically well isolated from all other quantum dots in the laser focus. Figure~\ref{fig:PL} shows the corresponding polarization resolved photoluminescence (PL) spectrum of the two quantum dots for two orthogonal linear polarization directions at a sample temperature of 5~K, a laser focus of 1~$\mu$m, and nonresonant excitation into the wetting layer by 5~$\mu$W linearly polarized light. The anisotropic exchange interaction leads to a splitting of the naturally linear polarized eigenstates for this type of QDs if they are uncharged. The two polarization components of the QD on the left side of the spectrum at 1.39607~eV do not show an anisotropic exchange interaction splitting which is a good indication for a positively charged quantum dot $(X^{+})$ resonance in our case. However, the QD at 1.39625~eV shows a pronounced splitting which indicates an uncharged QD resonance $(X^{0})$. Both cases are nicely confirmed by the SN measurements; i.e., the $X^{+}$ resonance shows spin noise, whereas the $X^{0}$ resonance does not contribute to the SN signal. The measured PL linewidth of the QD of $\approx 30\, \mu$eV is limited by the resolution of the spectrometer.

We carry out spin noise measurements on the exact same sample spot as the PL measurements by tuning the photon energy of an ultralow noise ring laser to the QD resonance. The photon energy is controlled by a high precision wavelength meter. The spin induced stochastic Faraday rotation of the reflected laser light is resolved outside the He dewar (see Ref.~\cite{Dahbashi2012} for further details of the experimental setup) by a polarization bridge featuring an extremely low noise balanced photo receiver with switchable bandwidth. The resulting electrical signal is amplified, digitized in the time domain, and Fourier transformed in real time. Remaining dc components are suppressed by cascaded highpass filters \cite{Note8}. The noise background due to optical shot noise of the laser and electrical noise of the balanced receiver and the amplifier is eliminated by subtracting spin noise spectra with longitudinal and transverse magnetic field from each other \cite{Note7}. This method works well since the applied transverse magnetic field drastically reduces the projection of the longitudinal spin component on the direction of detection. The inset in Fig.~\ref{fig:PL} shows a typical SN spectrum with a Lorentzian line shape, whereat the full width at half maximum $\nu_{\text{\tiny FWHM}}$ yields the longitudinal heavy hole spin relaxation rate $\Gamma_1 = 1/T_1 = \pi \nu_{\text{\tiny FWHM}}$. Magnetic fields of up to 31~mT are applied both in longitudinal and transverse geometry. All experiments are carried out at a fixed temperature of 5~K and focus on the $T_1$ spin noise which is centered at zero frequency. The SN power corresponding to $T_2^*$ does not influence the following magnetic field dependent measurements since the $T_2^*$ SN spectrum is much broader ($\approx 15$~MHz), lower in amplitude, even for $B_z=0$~mT not centered at zero frequency \cite{Glazov2012}, and efficiently suppressed due to the chosen bandwidth of the balanced receiver which is adapted to the $\Gamma_1$ SN linewidth and set for most of our experiments to 1.8~MHz \cite{Note1}.

\begin{figure}[tbp]
  \centering
  \includegraphics[width=0.99 \columnwidth]{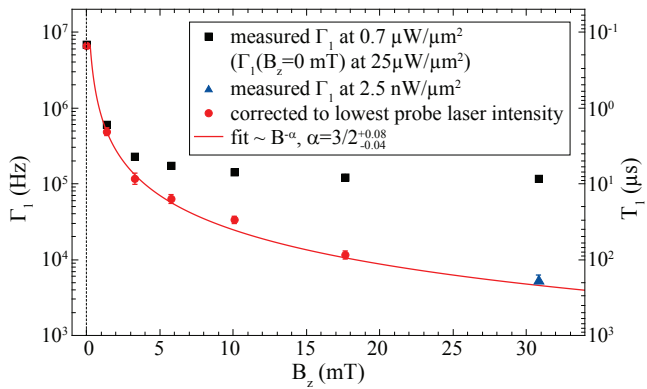}
  \caption{(Color online) Dependence of the spin relaxation rate of the Lorentzian spin noise spectrum on the longitudinal magnetic field. The black squares show the saturation of the linewidth at constant probe laser intensity due to optical QD excitation. The red dots depict the linewidth corrected to the lowest possible probe intensity. The red line is a fit to the intensity corrected linewidths with $B^{-\alpha}$ and $\alpha=3/2^{+0.08}_{-0.04}$.
  }
\label{fig:B_dependence}
\end{figure}

\begin{figure}[bp]
  \centering
  \includegraphics[width=0.95 \columnwidth]{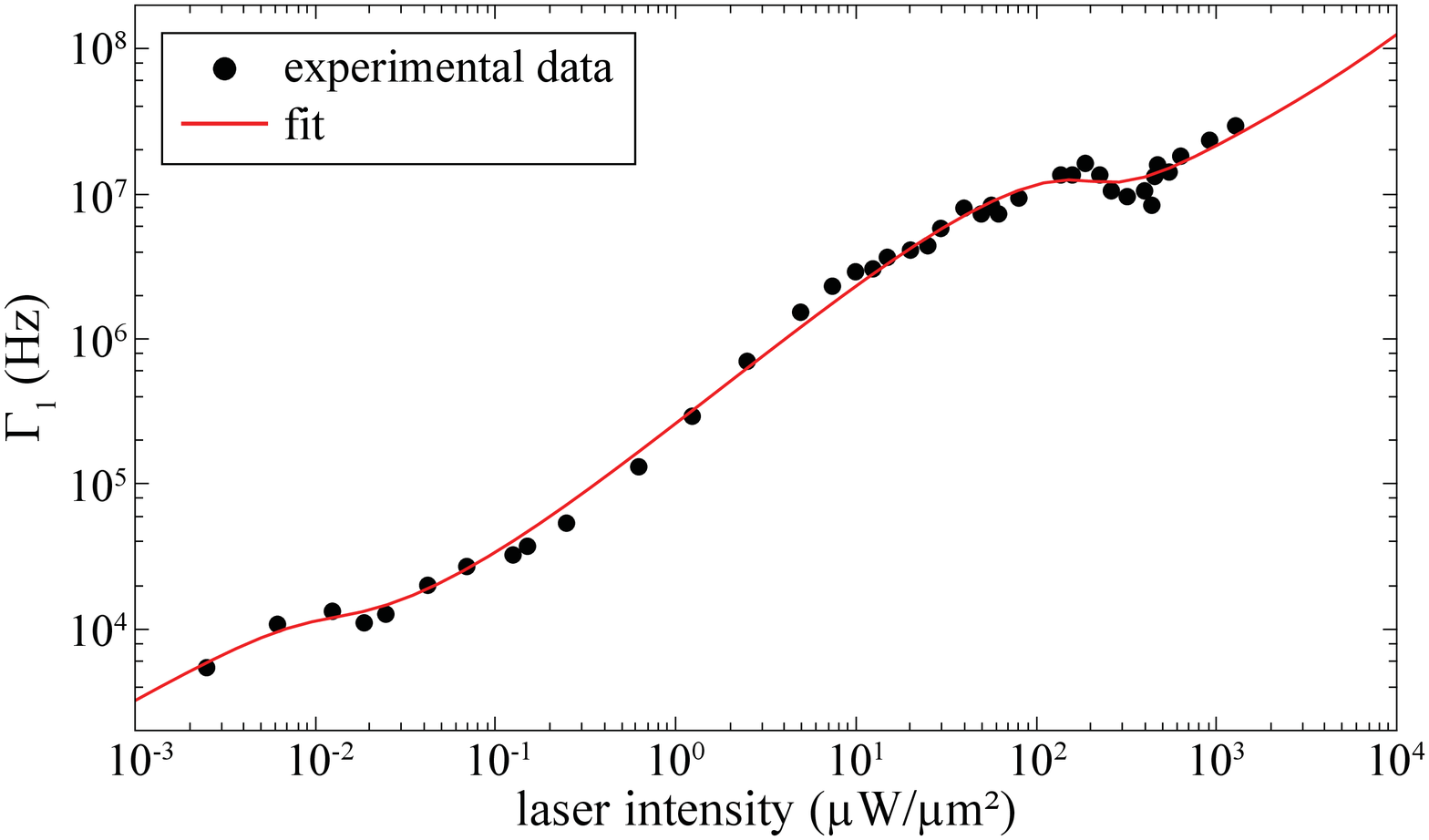}
  \caption{(Color online) Spin relaxation rate versus probe laser intensity at constant detuning. The deviation from a linear relation between $\Gamma_1$ and probe laser intensity reveals that the single QD has not a single resonance but experiences sustained step-like charge fluctuations in its vicinity.
 %Spin relaxation rate versus probe laser intensity at constant detuning. The deviation from a linear relation between $\Gamma_1$ and probe laser intensity reveals that the single QD has not a single resonance but experiences over time step-like charge fluctuations in its vicinity.
  The red line is a fit according to Eq.~\protect\ref{equ:noise} (see \cite{Note4} for the fitting parameters).
  %The red line is a calculation according to Eq.~\protect\ref{equ:noise} (see \cite{Note4} for the fitting parameters).
  }
\label{fig:Intensity}
\end{figure}

The blue squares in Fig.~\ref{fig:PL} depict the integrated spin noise power of the $\Gamma_1$-related SN contribution in dependence on the photon energy for a probe laser intensity of $0.7~\mu$W$/\mu$m$^2$.
%The blue squares in Fig.~\ref{fig:PL} depict the integrated $T_1$ spin noise power in dependence on the photon energy for a probe laser intensity of $0.7~\mu$W$/\mu$m$^2$.
The SN power spectrum coincidences very well with the PL spectrum of the charged QD and explicitly drops towards zero with increasing detuning from the resonance. These observations are unambiguous evidence for spin noise originating from one single quantum dot. We will discuss the exact shape of the SN power spectrum later and first focus on the magnetic field and intensity dependence of $\Gamma_1$. These measurements are carried out at the high energy slope of the SN power peak at an energy of $1.396075$~eV. The black squares in Fig.~\ref{fig:B_dependence} depict the measured magnetic field dependence of $\Gamma_1$ for a laser intensity of 0.7~$\mu$W$/\mu$m$^2$ \cite{Note2}. The logarithmic plot shows a strong decrease of $\Gamma_1$ with increasing magnetic field which starts to saturate above 10~mT. Such a saturation has been observed before by SNS in an ensemble of (InGa)As QDs (see Fig.~3 in Ref.~\cite{Li2012}). In order to identify the origin of this saturation, we measure the dependence of $\Gamma_1$ on laser intensity for 
% a longitudinal magnetic field 
$B_z=31$~mT. The black dots in Fig.~\ref{fig:Intensity} depict the measured intensity dependence of $\Gamma_1$ over 6 orders of magnitude. We will discuss the detailed structure of the intensity later but the measurement clearly shows a dramatic intensity dependence and proves that the afore observed saturation of $\Gamma_1$ results from laser excitation. In the case of laser excitation, the intrinsic spin relaxation rate is superimposed by the photon absorption rate since (a) a resonantly absorbed electron hole pair blocks the optical transition and (b) the second hole suppresses spin noise from the resident hole by Pauli blockade. We want to point out that laser induced broadening of the SN spectrum plays an especially complicated role in QD {\it ensembles} since the intensity broadened SN spectrum is nearly intensity independent over several orders of magnitude of laser intensity (see Refs.~\cite{Dahbashi2012,Oestreich2012} for details).

The blue triangle in Fig.~\ref{fig:B_dependence} depicts $\Gamma_1$ measured at 2.5~nW$/\mu$m$^2$ and $B_z=31$~mT from Fig.~\ref{fig:Intensity}. 
% A comparison with $\Gamma_1$ at 0.7~$\mu$W$/\mu$m$^2$ and $B_z=31$~mT (black square) shows that the $\Gamma_1$ measurement at 0.7~$\mu$W$/\mu$m$^2$ is to a very large extent dominated by the laser intensity. 
In the following we correct the data measured between $0 < B_z < 31$~mT and 0.7~$\mu$W$/\mu$m$^2$ by subtracting the difference of the rates measured at $B_z=31$~mT at high and low laser intensity \cite{Note9}. The results are shown as red dots in Fig.~\ref{fig:B_dependence} and represent an upper limit for the intrinsic $\Gamma_1$ since they still contain a small extrinsic contribution which results from the finite laser intensity of 2.5~nW$/\mu$m$^2$. Fortunately, this contribution is in any case smaller than the rate depicted by the blue triangle and in good approximation negligible between $B_z=0$ and 10~mT where the corrected data yield a strong magnetic field dependence of $B^{-\alpha}$ with $\alpha=3/2^{+0.08}_{-0.04}$. The extrapolation of the fit to 31~mT indicates also for $B_z>10$~mT a good agreement with the $B^{-\alpha}$ dependence.

% The red dots in Fig.~\ref{fig:B_dependence} depict the measured $\Gamma_1$ corrected by the photon absorption rate which has been extracted from the $B_z=31$~mT measurement at 0.7~$\mu$W$/\mu$m$^2$ where the intrinsic spin relaxation rate is \tcb{in good approximation} negligible compared to the photon absorption rate. The depicted red dot at $B_z=31$~mT has been measured at the currently lowest possible laser power for SN measurements of 2.5~nW and is according to Fig.~\ref{fig:Intensity} a \tcr{present upper bound for $\Gam ma_1$}. \tcr{The red dots at lower magnetic fields are corrected by this constant value of the influence by probe light absorption on $\Gamma_1$}. The solid red line in Fig.~\ref{fig:B_dependence} is a fit to the \tcr{corrected, i.e., the present upper limit of the intrinsic,} heavy hole spin relaxation rate \tcr{$\Gamma_1$} which yields a \tcr{strong} magnetic field dependence of $B^{-\alpha}$ with $\alpha=3/2^{+0.08}_{-0.04}$ \tcr{as a first evidence}.
%The depicted red dot at $B_z=31$~mT has been measured at the currently lowest possible laser power for SN measurements of 2.5~nW and is according to Fig.~\ref{fig:Intensity} a lower bound for $T_1$. The solid red line in Fig.~\ref{fig:B_dependence} is a fit to the intrinsic heavy hole spin relaxation rate which yields a magnetic field dependence of $B^{-\alpha}$ with $\alpha=3/2^{+0.08}_{-0.04}$.

%\footnote{The absorption probability of the single QD on resonance $\alpha_0=5\cdot10^{-6}$. The intrinsic spin relaxation rate at $B_z=31$~mT at zero intensity $\gamma_{I=0}=100$~Hz. The detuning of the three QD configurations in units of the QD homogeneous radiative linewidth are 2.6, 11.9, and 55.3. The respective quantifiers $G_i$ are 1.08, 0.89, and 1.12.}

% This explicit 
The strong magnetic field dependence reflects a part of the very intricate hole spin and nuclear spin bath interaction most commonly known as central spin problem where a single (central) spin is influenced by a spin bath of nuclear spins which are in first approximation totally random \cite{Gaudin.J.Phys.France.1976}. However, mutual interaction of the central spin and the spin bath acts back on the spin dynamic of the central spin due to  the Knight field and results in a slow but finite spin dephasing \cite{PhysRevB.81.115107, Urbaszek.RMP.2013}. There exists a plethora of theoretical works addressing this problem in solid state, quantum optics, and chemistry systems predicting for example exponential decays \cite{PhysRevB.77.125329} or $1/\rm{log}(t)$ like behavior \cite{AlHassanieh.PRL.2006} for the explicit temporal evolution of the spin correlator. A straightforward experimental method to study this dynamic is the application of an external magnetic field which adds to the random nuclear field and thus alters the spin back action. For intermediate magnetic fields up to a few Tesla Trif \textit{et al.} \cite{Trif2009} predict a rising spin relaxation rate $\Gamma_1$ with $B_z$ due to phonon induced spin flips, whereas Sinitsyn \textit{et al.} \cite{Sinitsyn2012} calculate approximately $\Gamma_1 \propto e^{-B}$ due to nuclear quadrupole coupling (see Fig.~4 in Ref. \cite{Sinitsyn2012}). On the other hand Fras \textit{et al.} \cite{Fras2012a} extract $\Gamma_1 \propto O(B^{-2})$ from their measurements which saturates at higher fields. However, here we observe so far no saturation of the decreasing spin relaxation rate with rising magnetic field and expect that the temporal dynamic of the longitudinal spin component will slow down even further with increasing field. %For the exact theoretical value of the exponent $\alpha$ one has to consider furthermore the shape of the hole wavefunction and the value for the Zeeman splitting anisotropy in more detail, which is difficult to extract for this very quantum dot in the presented measurements. 
The exact theoretical value of the exponent $\alpha$ needs more sophisticated calculations which should include the shape of the hole wave function and the value for the Zeeman splitting anisotropy. 
%Additionally we want to point out that we observe a steadily increasing integrated noise power with increasing magnetic field, which we attribute to a continuous redistribution of spin noise into the longitudinal spin noise signal (not shown). \tcg{see supplemental: to be discussed!}

Next, we want to discuss the line shape of the integrated SN power \cite{Note6} in Fig.~\ref{fig:PL} and the intricate intensity dependence of $\Gamma_1$ in Fig.~\ref{fig:Intensity}. One might expect for a single QD a SN spectrum with two sharp maxima which corresponds to the square of the imaginary part of the refractive index of a single Lorentzian absorption line with a homogeneous linewidth of about 1.5~$\mu$eV for a typical (InGa)As QD. However, the measured SN power in Fig.~\ref{fig:PL} yields a single Gaussian like peak with a full width at half maximum of 19~$\mu$eV \cite{Note3}. We attribute the origin of this Gaussian like peak to single charge fluctuations in the local vicinity of the QD. These charge fluctuations always occur since we do not use the usual \textit{pin}-structure \cite{atature2007, berezovsky2006, degreve2011} to avoid any influences of an built-in electric field on the spin relaxation time. Such local charge fluctuations are common in semiconductor physics \cite{Efros.PRL.1997} and arise for example in diamond NV centers \cite{Tamarat.PRL.2006, Sipahigil.PRL.2012} and embedded nanocrystals \cite{Empedocles.PRL.1996, Nirmal.Nature.1996}. Houel et al. \cite{Houel2012} showed very recently that single charge fluctuations of a small number of defects located within $\sim$100~nm of an molecular beam epitaxy grown (InGa)As QD yield due to the single charge induced Stark shift between 3 and 6 step-like shifts of the QD resonance with a total shift in energy of typically 30~$\mu$eV. This stochastic energy shift is in good agreement with the Gaussian width of the SN peak. These charge fluctuations also explain the intricate intensity dependence in Fig.~\ref{fig:Intensity}. 
The red solid line in Fig.~\ref{fig:Intensity} depicts a fit based on a numerical model of the intensity dependence of $\Gamma_1$ assuming three different charged defect configurations in the vicinity of the QD. 
%The red solid line in Fig.~\ref{fig:Intensity} depicts calculations of the intensity dependence of $\Gamma_1$ assuming three different charge configurations in the vicinity of the QD.
The underlying calculations are an extension of the SN QD ensemble model of Refs.~\cite{Dahbashi2012,Oestreich2012} to a single QD with local charge variations.
%The calculations are an extension of the SN QD ensemble model of Ref.~\cite{Dahbashi2012,Oestreich2012} to a single QD with local charge variations.
The intensity dependent spin noise power spectrum results from the sum over all QD configurations with the quantifier $G_i$, which specifies the probability of each configuration $i$ averaged over time:
\begin{eqnarray}\label{equ:noise}
P_{\text{\tiny SN}}(\nu_s) = \sum_{i} &G_i&\times \left(n (\Delta E_i)\right)^2\times  \\
&R^2&(\Delta E_i,I)\times L\left(\gamma(\Delta E_i,I),\nu_s\right), \nonumber
\end{eqnarray}
where $n$ is the dispersive part of the refractive index, $L$ the Lorentzian SN spectrum centered at zero frequency, and $\Delta E_i$ the relative detuning of the probe photon energy with respect to the resonance energy of the specific QD configuration $i$. Interestingly all three quantifiers, which result from the fit to the experimental data, deviate only weakly, i.e., all three charge configurations are nearly equally probable which is in good agreement with Ref. \cite{Houel2012}. The reduction of the SN power due to excitation-induced Pauli blockade is accounted for by
\begin{equation}\label{equ:reduction}
  R(\Delta E_i,I)=1/(1+\tau_{\rm PL}/\tau_{\text{\tiny free}}(\Delta E_i,I)), 
\end{equation}
where $\tau_{\rm PL}$ is the radiative lifetime of the QD and
\begin{equation}\label{equ:frei}
  \tau_{\text{\tiny free}} (\Delta E_i,I)=1/(\gamma_{ex}(I) \alpha_{0}(\Delta E_i))
\end{equation}
the time between two excitations, where $\gamma_{ex}$ is the number of photons from the laser per second and $\alpha_{0}$ the Lorentz-shaped QD absorption. The line shape of the calculated SN spectrum is in first approximation also Lorentzian like since one QD configuration usually dominates the SN spectrum \cite{Note5}. This is in good agreement with the experimental observations and we define the spin relaxation rate $\Gamma_1$ as $\pi \nu_{\text{\tiny FWHM}}$.

The accordance of the experimental data and the fit based on the numerical model shows that the intricate intensity dependence of $\Gamma_1$ results from the detuning dependent consecutive broadening of the SN spectrum of each discrete QD configurations. 
%The calculations show that the intricate intensity dependence of $\Gamma_1$ results from the detuning dependent consecutive broadening of the SN spectrum of each discrete QD configurations.
At extremely low laser intensities, the photon absorption rate is smaller than $\Gamma_1$ and the width of the SN spectrum is dominated by the intrinsic spin relaxation rate. This condition is probably still not yet entirely satisfied in Fig.~\ref{fig:Intensity} despite a minimum laser intensity of only 2.5~nW. With increasing intensity, the QD configuration with the smallest detuning from the laser starts to broaden significantly which yields an increase of the measured $\Gamma_1$ since this QD contributes most significantly to the SN spectrum. 
%With increasing intensity, the QD configuration with the smallest detuning from the laser starts to broaden significantly which yields an increase of the measured $\nu_{\text{\tiny FWHM}}$ since this QD contributes most significantly to the SN spectrum.
However, a further increase of laser intensity broadens the most resonant QD configuration so much that its SN amplitude falls below the amplitude from the other QD configurations which are less broadened.
%However, a further increase of laser intensity broadens the most resonant QD configuration to such an extend that its SN amplitude becomes lower than the amplitude from the other QD configurations which are less broadened.
This scenario takes place consecutively for the other QD configurations. As a consequence, the measured $\Gamma_1$ increases at some intensities only sub-linear or even drops (see Fig.~\ref{fig:Intensity} at around 20~nW$/\mu$m$^2$ and 300~$\mu$W$/\mu$m$^2$). 
%\tcg{Diesen Satz raus: oder am Ende Ref zu OestreichSPIE2012: }At very high laser intensities, the photon absorption rate is not negligible any more in comparison to $\tau_{\text{\tiny PL}}^{-1}$ and also the integrated SN power decreases (not shown).
%As a consequence, the measured $\nu_{\text{\tiny FWHM}}$ increases at some intensities only sub-linear or even drops (see Fig.~\ref{fig:Intensity} at around 300~$\mu$W$/\mu$m$^2$). At very high laser intensities, the photon absorption rate is not negligible any more in comparison to $\tau_{PL}^{-1}$ and also the integrated SN power decreases (not shown).

In summary, spin noise spectroscopy has reached the ultimate level of single spin detection and shows a dramatic magnetic field dependence of the heavy hole spin relaxation rate at low magnetic fields. The combination of magnetic field and intensity dependent measurements on a single QD reveals - in contrast to other experiments - that $T_1$ does not saturate but increases by three orders of magnitude from 0~mT to 31~mT. 
Such an increase excludes several theories concerning the heavy hole spin relaxation in (InGa)As quantum dots but is in qualitative agreement with recently published calculations of the relaxation of central spins which include the nuclear quadrupole coupling.
%Such an increase excludes several theories concerning the heavy hole spin relaxation in (InGa)As quantum dots but is in qualitative agreement with recent calculations of the relaxation of central spins which include the nuclear quadrupole coupling.
Additionally, the measured linewidth of the integrated spin noise power versus laser energy is significantly broader than the transform limited optical linewidth of a single (InGa)As quantum dot. This effect is attributed to single charge fluctuations in the quantum dot vicinity. The charge fluctuations also manifest themselves in the intricate intensity dependence of the measured width of the spin noise spectrum and will be important in the context of quantum dot entanglement, spin qubits, and spin-orbit-mediated manipulation of heavy hole spins in semiconductor nanodevices.

We acknowledge the financial support by the BMBF joint research project QuaHL-Rep, the Deutsche Forschungsgemeinschaft in the framework of the priority program ``SPP 1285-Semiconductor Spintronics,'' and the excellence cluster ``QUEST-Center for Quantum Engineering and Space-Time Research''.

%\bibliographystyle{apsrev4-1}   % pr-jens / unsrt
%\bibliography{library,single_dot_sns_001}


%

\end{document}